\documentclass[prd,aps,tightenlines,superscriptaddress,amsmath,amssymb,amsfonts,twocolumn]{revtex4-2}
\usepackage{amssymb,amsmath,amsthm,amsbsy,epsfig,color,graphicx,times}
\usepackage[ansinew]{inputenc}
\usepackage[english]{babel}
\usepackage{color}
\usepackage{braket}
\usepackage{tabularx}
\usepackage{array}

\begin{document}
\title{Neutrinos, mixed bosons, Quantum Reference Frames and entanglement}

\author{A. Capolupo}
\email{capolupo@sa.infn.it}
\affiliation{Dipartimento di Fisica ``E.R. Caianiello'' Universit\`{a} di Salerno, and INFN -- Gruppo Collegato di Salerno, Via Giovanni Paolo II, 132, 84084 Fisciano (SA), Italy}

\author{A. Quaranta}
\email{anquaranta@unisa.it}
\affiliation{Dipartimento di Fisica ``E.R. Caianiello'' Universit\`{a} di Salerno, and INFN -- Gruppo Collegato di Salerno, Via Giovanni Paolo II, 132, 84084 Fisciano (SA), Italy}

\begin{abstract}

We discuss the relevance of quantum reference frames in the description of mixed particle states. We show that the notion of rest frame for mixed particles, which is classically ill-defined, can be introduced in the context of quantum frames. We discuss the possible phenomenological implications, displaying a new form of frame-dependent entanglement that characterizes reactions involving mixed particles.

\end{abstract}

\maketitle

\textbf{Introduction} -- The principle of relativity is one of the very foundations of physics. The equivalence of physical laws in all the admissable reference frames is a basic requirement for any sensible theory. The notion of reference frames, and the classes thereof on which the requirement of form invariance of the physical laws is to be demanded, have changed radically in time. Special relativity has updated the notion of inertial frames by the inclusion of clocks, while general relativity has enlarged the class of equivalent frames by including non-inertial frames and by the requirement of general covariance. But ultimately a reference frame is an abstraction for a reference physical system. Since physical systems are believed to be fundamentally quantum, a generalized notion of \emph{quantum reference frames} has to be introduced.
Historically quantum reference frames have first emerged in quantum information \cite{QI1,QI2,QI3,QI4,QI5,QIR,QI6,QI7,QI8,QI9,QI10,QI11,QI12}, but the related issue of relational observables has been discussed extensively in Quantum Gravity \cite{QG1,QG2,QG3,QG4}, where the notion of quantum reference frame may play an important role. Along the relational paradigm, advocated for instance in Refs. \cite{QIR,QIR2}, in the last years a foundational and relational approach to quantum reference frames has been developed \cite{FQR1,FQR2,FQR3,FQR4,FQR5,FQR6}, together with a notion of quantum transformations between reference frames \cite{FQR1,FQR2,FQR4}.

In this letter we wish to discuss the relevance of quantum reference frames in particle physics, in particular with respect to mixed particles such as neutrinos and neutral mesons. The fundamental role of the symmetry group of transformations among reference frames was elucidated by Wigner \cite{Wigner}. Specifically particles are strictly related to the representations of the Poincaré group of special relativity. We here show that mixed particle states require quantum reference frames to be properly interpreted. We show that the ``rest frame'' of a mixed particle, indeed, can only arise through a (strictly) quantum transformation of frames. Among the peculiarities of quantum frames we discuss the relativity of entanglement, and how the latter may characterize reactions involving mixed particles.
Our analysis hints at the necessity of generalizing the work of Ref.\cite{Wigner} in the sense of quantum frame transformations in order to account for mixed elementary particles (neutrinos).

\textbf{Flavor neutrino states} -- The seminal paper by Wigner \cite{Wigner} achieved the complete classification of the projective representations of the Poincaré group. One particle states of relativistic quantum field theory belong to any of the representations $\ket{m,s}$ distinguished by mass $m$ and spin $s$, which are essentially related to the Casimir operators of the Poincaré group. Tipically one employs the momentum basis $\ket{p^{\mu}, \sigma }_{m,s}$ of the simultaneous eigenstates of the translation operators $P^{\mu}$, with $P^{\mu} \ket{p^{\mu}, \sigma }_{m,s} = p^{\mu} \ket{p^{\mu}, \sigma }_{m,s}$, on which the Poincaré group acts unitarily as \cite{Wigner,Weinberg} $U (\Lambda) \ket{p^{\mu}, \sigma }_{m,s} = \sum_{\sigma'} D^{(s)}_{\sigma' \sigma} (W(\Lambda, p)) \ket{\Lambda^{\mu}_{\nu} p^{\nu}, \sigma' }_{m,s} $  and $U(a) \ket{p^{\mu}, \sigma }_{m,s} = e^{-i p^{\mu} a_{\mu}} \ket{p^{\mu}, \sigma }_{m,s}$. Here $\Lambda$ labels the homogeneous Lorentz transformations, $a^{\mu}$ is the translation $4$-vector, the coefficients $D^{s}_{\sigma' \sigma} (R)$ implement the $2s + 1$-dimensional unitary irreducible representation of the rotation group (we are assuming $m > 0$), and $W(\Lambda,p)$ is the Wigner rotation. Elementary particles are then understood as belonging to the unitary irreducible representations of the Poincaré group with a given mass $m$ and a given spin $s$.

There is one notable exception to this scheme. On the one hand flavor neutrinos \cite{Neut1,Neut2,Neut3,Neut4,Neut5} are considered as elementary particles, on the other hand they do not belong, strictly speaking, to a unitary irreducible representation of the Poincaré group. Taking the Pontecorvo states literally, the electron neutrino state is given by the superposition
\begin{equation}\label{NeutrinoState}
 \ket{\pmb{p}, e} = \cos \theta \ket{p^{\mu}_{1}}_{m_1} + \sin \theta \ket{p_{2}^{\mu}}_{m_2}
\end{equation}
where we ignore spin and consider only two flavors for simplicity. The momenta are specified as $p_{j}^{\mu} \equiv \left( \pmb{p}, \omega_j \right)$ with the on-shell energies $\omega_j = \sqrt{\pmb{p}^2 + m_j^2}$. Clearly the states of Eq. \eqref{NeutrinoState} are eigenstates of the $3$-momentum $\pmb{P} \ket{\pmb{p},e} = \pmb{p} \ket{\pmb{p},e}$, but not eigenstates of the energy $P^0$. A similar expression holds for the muon neutrino state. In each case flavor neutrinos involve two distinct mass hyperboloids and two distinct irreducible representations $m_1, m_2$. While the notion of elementary particle is more a matter of nomenclature, the peculiar superposition of Eq. \eqref{NeutrinoState} has a concrete impact e.g. on the weak interaction processes. Consider for instance the $\beta$ decay $n \rightarrow p + e^{-} + \nu_e$. The total $4$-momentum conservation for such a process cannot be enforced unless at least one of the other particles $n, p, e$ carries a $4$-momentum uncertainty (specifically an energy uncertainty, for the state of Eq. \eqref{NeutrinoState}) which is able to compensate for the
lack of definiteness in the $4$-momentum of $\nu_e$. This kind of 'essential' uncertainty is instead not required for all those processes which involve only on-shell asymptotic states (e.g. $e^{-} + e^{+} \rightarrow 2 \gamma$).

\textbf{The rest frame of flavor neutrinos} -- For particles with well defined $4$-momentum $\ket{p^{\mu}}_{m}$ there is a natural notion of 'rest frame', which describes the physics as 'sitting on the particle'. In its rest frame the particle has the state $\ket{p^{\mu}_{(0)}}_{m} \equiv \ket{(\pmb{0},m) }_{m}$ corresponding to vanishing $3$-momentum. There is a unique Lorentz transformation $\Lambda$, up to spatial rotations, that takes the particle from the frame in which it has momentum $p^{\mu}$ to its rest frame, where it has momentum $p^{\mu}_{(0)}$, namely

\begin{equation}
 p^{\mu}_{{0}} = \Lambda^{\mu}_{\nu} p^{\nu} \ .
\end{equation}

Assume, without loss of generality, that in a given frame the particle has $4$-momentum $p^{\mu} \equiv \left(\sqrt{p^2 + m^2},0,0,p \right)$ (if this is not the case, just perform a rotation to bring the momentum along the $z$ axis). Then the transformation to the rest frame is performed by the boost

\begin{equation}\label{ClassicalTransformation}
 \Lambda^{\mu}_{\nu} \equiv \begin{pmatrix} \frac{\sqrt{p^2 + m^2}}{m} & 0 & 0 & \frac{-p}{m} \\ 0 & 1 & 0 & 0 \\ 0 & 0 & 1 & 0 \\ \frac{-p}{m} & 0 & 0 & \frac{\sqrt{p^2 + m^2}}{m} \end{pmatrix} \ .
\end{equation}

Notice that the transformation is controlled by the $c$-number parameter $\psi = \cosh^{-1} \left( \frac{\sqrt{p^2 + m^2}}{m} \right)$. What is the equivalent notion of rest frame for a flavor neutrino with state of Eq. \eqref{NeutrinoState}? We want the neutrino state to have a vanishing $3$-momentum in its own rest frame, namely

\begin{equation}\label{NeutrinoRestState}
 \ket{\pmb{0}, e} = \cos \theta \ket{\left(\pmb{0}, m_1 \right),e}_{m_1} + \sin \theta \ket{\left(\pmb{0}, m_2 \right),e}_{m_2}
\end{equation}

where we have written out explicitly the momentum components. It is easy to recognize that \emph{there is no classical Lorentz transformation (Eq. \eqref{ClassicalTransformation}), regulated by a $c$-number boost parameter, which is able to transform the state of Eq. \eqref{NeutrinoState} into the state of Eq. \eqref{NeutrinoRestState}}. To see this, let us, for instance, try to annihilate the $3$-momentum $\pmb{p}$ of the $m_1$ component $\ket{\left(\pmb{p}, \sqrt{\pmb{p}^2 + m_1^2} \right)}_{m_1}$. Assuming, without loss of generality, the $3$-momentum along the $z$ axis, we perform a Lorentz boost $\Lambda_1$ of the form Eq. \eqref{ClassicalTransformation}, with boost parameter $\psi_1 = \cosh^{-1} \left(\frac{\sqrt{p^2 + m_1^2}}{m_1} \right)$. We then have

\begin{equation}\label{TransformedState}
 U (\Lambda_1) \ket{\pmb{p},e} = \cos \theta \ket{\left( \pmb{0}, m_1 \right) }_{m_1} + \sin \theta \ket{\left( \tilde {\pmb{p}}, \tilde{E} \right) }_{m_2}
\end{equation}

with $\tilde{\pmb{p}} \equiv \left(0,0, \frac{p \left(\sqrt{p^2 + m_1^2} - \sqrt{p^2 + m_2^2}\right)}{m_1} \right)$ and $\tilde{E} = \frac{\sqrt{\left(p^2 + m_1^2 \right) \left(p^2 + m_2^2 \right)} - p^2}{m_1}$. Evidently the state Eq. \eqref{TransformedState} is \emph{not} the rest state that we sought (Eq. \eqref{NeutrinoRestState}). More generally, a classical Lorentz transformation should have a boost parameter $\psi$ such that $\sqrt{p^2 + m_1^2} \sinh \psi + p \cosh \psi = 0 = \sqrt{p^2 + m_2^2} \sinh \psi + p \cosh \psi$ in order to transform the state of Eq. \eqref{NeutrinoState} into the state of Eq. \eqref{NeutrinoRestState}. This is impossible, since $\sqrt{p^2 + m_1^2} \neq \sqrt{p^2 + m_2^2}$. Albeit no classical Lorentz transformation can bring the state of Eq. \eqref{NeutrinoState} to the state of Eq. \eqref{NeutrinoRestState}, a \emph{quantum} Lorentz transformation can do the trick. By this we mean that an appropriate superposition of Lorentz boosts can indeed perform the transformation to the 'rest frame' of the neutrino. Consider the Hilbert spaces $\mathcal{H}_{m_j}$ spanned by the one particle states of mass $m_j$ and the projectors $\mathcal{P}_j = \int d^4 p \ \theta (p^0) \ \delta (p^{\mu}p_{\mu}-m_j^2)  \ket{p^{\mu}}_{m_j} \bra{p^{\mu}}_{m_j}$ acting naturally on the full one particle sector $ \mathcal{H} = \bigoplus_{m > 0} \mathcal{H}_{m}$. On the subpace $\mathcal{H}_{m_j}$ the projector $\mathcal{P}_j$ is simply the identity, while $\mathcal{P}_j$ annihilates all the states with masses $m \neq m_j$. The transformation to the 'rest frame' of the neutrino can then be realized by the operator
\begin{equation}
 \mathcal{U} = U(\Lambda_1) \mathcal{P}_1 +  U(\Lambda_2) \mathcal{P}_2 \ .
\end{equation}
The Lorentz boosts $\Lambda_j$ have the form of Eq. \eqref{ClassicalTransformation} for the mass $m_j$. More simply we can promote the classical Lorentz boost of Eq. \eqref{ClassicalTransformation} to an operator. Here it is the boost parameter itself $\psi$, which is turned into an operator $\hat{\psi} = \cosh^{-1} \left( \hat{H}\left( \hat{P}^{\mu} \hat{P}_{\mu} \right)^{-\frac{1}{2}} \right)$, with $\hat{H} = \hat{P}^0$, and the hats were used to remark that these quantities are operators on $\mathcal{H}$. The quantum Lorentz boost has then the form
\begin{equation}\label{QuantumTransform}
  \hat{\Lambda}^{\mu}_{\nu} \equiv \begin{pmatrix}  \hat{H}\left( \hat{P}^{\mu} \hat{P}_{\mu} \right)^{-\frac{1}{2}} & 0 & 0 & -\hat{P}_z\left( \hat{P}^{\mu} \hat{P}_{\mu} \right)^{-\frac{1}{2}} \\ 0 & 1 & 0 & 0 \\ 0 & 0 & 1 & 0 \\ -\hat{P}_z\left( \hat{P}^{\mu} \hat{P}_{\mu} \right)^{-\frac{1}{2}} & 0 & 0 &  \hat{H}\left( \hat{P}^{\mu} \hat{P}_{\mu} \right)^{-\frac{1}{2}} \end{pmatrix}  \
\end{equation}
The corresponding unitary representation $U(\hat{\Lambda})$ transforms the neutrino state of Eq. \eqref{NeutrinoState} into its rest form (Eq. \eqref{NeutrinoRestState}). The action of $U(\hat{\Lambda})$ clearly reduces to that associated to an ordinary Lorentz boost when acting on states with definite $4$-momentum:

\begin{equation}\label{QuantumTransform2}
 U ( \hat{\Lambda} ) \ket{p^{\mu}}_m = U(\Lambda (p) ) \ket{p^{\mu}}_{m} = \ket{\Lambda^{\mu}_{\nu} (p) p^{\nu}}_{m}
\end{equation}

with $\Lambda^{\mu}_{\nu} (p)$ provided by Eq. \eqref{ClassicalTransformation}. Equation \eqref{QuantumTransform} also defines the action of $U (\hat{\Lambda})$ on a generic element of $\mathcal{H}$ by linearity.

The quantum nature of the transformation reveals that the 'rest frame' of a flavor neutrino cannot be understood as a classical reference frame, but only as a quantum reference frame. Transformations like that of Eq. \eqref{QuantumTransform}, in which the parameter of the transformation itself is an operator, do indeed represent the kind of generalized transformations that occur in transforming between quantum reference frames \cite{FQR2,FQR4}. This suggests that quantum reference frames are needed to properly interpret the one particle state corresponding to a flavor neutrino. Notice that the need for a quantum transformation of the neutrino state also characterizes other frames. For instance, moving to a frame where the neutrino has a definite $3$-momentum $\pmb{p}'$, from a frame where it has momentum $\pmb{p}$, while also preserving the equal $3$-momentum superposition of Eq. \eqref{NeutrinoState}, requires that one employs a quantum Lorentz boost with parameter $\hat {\psi} = \sinh^{-1} \left( \left(\hat{P}^{\mu} \hat{P}_{\mu}\right)^{-1} \left(p' \hat{H} - \sqrt{p^{'2} + \hat{P}^{\mu} \hat{P}_{\mu}} \hat{P}_z \right) \right)$. Here we have assumed, without loss of generality, that both $\pmb{p}$ and $\pmb{p}'$ lie along the $z$ axis. The transformation to the rest frame is just the special case $p' = 0$.

It can be rightfully argued that a quantum Lorentz transformation is needed to define the 'rest frame' of a generic particle with indefinite $4$-momentum, such as a wave packet $\ket{\phi} = \int d^4 p \ \theta (p^0) \ \delta (p^{\mu} p_{\mu} - m^2) \phi (p) \ket{p^{\mu}}_{m}$. This is indeed the case, and quantum transformations are required whenever momentum superpositions occur, as shown, for instance, in \cite{FQR2}. Yet the neutrino flavor states are different from the usual wave packets in that quantum transformations are \emph{fundamentally} involved in defining the related rest frame. To begin with, while wave packet are usually constructed superposing states that belong to the same mass hyperboloid, flavor states require at least two distinct masses to be defined. Secondly, there is no \emph{fundamental} reason for which an elementary particle cannot be in a sharp $4$-momentum eigenstate. On the contrary flavor states are necessarily in a superposition of $4$-momenta and thus carry an essential uncertainty.

\textbf{Relativity of entanglement} -- One of the striking features of quantum reference frames is that entanglement becomes a relative concept: it may be present in some frames while absent in others. Consider for simplicity three quantum particles $A, B, C$ on the line and assume that the state of $B$ and $C$, relative to A, is $\psi_{BC}^{(A)} = \ket{x_B}_B^{(A)} \ket{x_C}_C^{(A)}$. Here the notation is borrowed from \cite{FQR2,FQR5}, where the lower index is used to denote the quantum system and the upper index labels the quantum reference frame. The state of $A$ relative to $A$, $\ket{x_A = 0}_A$ is usually omitted \cite{FQR2,FQR5} since only relational information is retained. In order to shift to the point of view of $C$, one simply translates by the position $-x_C$ of $C$. The state of $A$ and $B$ relative to $C$ reads $\psi_{AB}^{(C)} = T(-x_C) \left( \ket{x_A = 0}_A \ket{x_B}_B \right) = \ket{-x_C}_A^{(C)} \ket{x_B - x_C}_B^{(C)}$. Here $T(a)$ is the unitary operator that performs the translation by $a$ on the states. Suppose now that the state of $B$ and $C$ relative to $A$ has the form $\psi^{(A)}_{BC} = \ket{x_B}_B^{(A)} \left(\ket{x_C^1}_C^{(A)} + \ket{x_C^2}_C^{(A)} \right)$. To shift to the point of view of $C$, we again translate by the position of $C$. But in this case the position of $C$ is not sharply defined, and, respecting linearity, we obtain the state of $A$ and $B$ relative to $C$ as
\begin{eqnarray}\label{EntangledState}
 \nonumber && \psi^{(C)}_{AB} = \left(T(-x^1_C) + T(-x^2_C) \right) \left(\ket{x_A = 0}_A \ket{x_B}_B \right) \\
 && = \ket{-x^1_C}_A^{(C)} \ket{x_B -x^1_C}_B^{(C)}  +  \ket{-x^2_C}_A^{(C)} \ket{x_B -x^2_C}_B^{(C)} \ .
\end{eqnarray}
The state $\psi^{(C)}_{AB}$ of Eq. \eqref{EntangledState} is clearly entangled, whereas the original state $\psi_{BC}^{(A)}$ was factorizable. The translation appearing in Eq. \eqref{EntangledState} is a quantum transformation in which the translation parameter, the position $x_C$ of $C$, is effectively turned into an operator $\hat{x}_C$.
The situation is similar for relativistic particles, except that we consider the $4$-momenta as the relevant degrees of freedom and the elements of the Poincaré group, instead of the translations on the line, as those performing the shift between classical reference frames. Suppose that there are three particles $A,B, \nu$ and that the state of $B$ and $\nu$, relative to $A$, is given by
\begin{equation}
 \psi_{B\nu}^{(A)} = \ket{p_B^{\mu}}_B^{(A)} \left( \cos \theta \ket{p^{\mu}_{1}}_{m_1, \nu}^{(A)} + \sin \theta \ket{p_{2}^{\mu}}^{(A)}_{m_2, \nu} \right) \ ,
\end{equation}
that is, the state of the neutrino $\nu$ relative to $A$ is that of Eq. \eqref{NeutrinoState}. Of course, $A$ is in its rest frame, so that its state reads $\ket{p^{\mu}_{A,0}}_A^{(A)} = \ket{(\pmb{0},m_A)}_A^{(A)}$. We assume, without loss of generality, that the spatial momentum $\pmb{p}$ of $\nu$ is along the $z$ axis. Shifting to the point of view of $\nu$ is tantamount to transform to the neutrino rest frame via the quantum transformation defined by Eqs. \eqref{QuantumTransform} and \eqref{QuantumTransform2}. The state of $A$ and $B$ becomes
\begin{eqnarray}\label{NeutrinoEntangledState}
 \nonumber  \psi_{AB}^{(\nu)} &=& \left(\cos \theta  \ U(\Lambda_1) + \sin \theta \ U(\Lambda_2) \right) \left( \ket{p^{\mu}_{A,0}}_A \ket{p^{\mu}_B}_B\right)  \\ \nonumber
 &=&  \cos \theta \ket{\Lambda^{\mu}_{1,\nu}p_{A,0}^{\nu}}_A^{(\nu)} \ket{\Lambda^{\mu}_{1,\nu}p_{B}^{\nu}}_B^{(\nu)} \\
 &+& \sin \theta \ket{\Lambda^{\mu}_{2,\nu}p_{A,0}^{\nu}}_A^{(\nu)} \ket{\Lambda^{\mu}_{2,\nu}p_{B}^{\nu}}_B^{(\nu)} \ .
\end{eqnarray}
Here it is understood that the operators $U(\Lambda)$ act on the states of $A$ and $B$, and that $\Lambda_j$ are the (classical) Lorentz boosts of Eq. \eqref{ClassicalTransformation} corresponding to masses $m_1$ and $m_2$. Notice the similarity between Eq. \eqref{NeutrinoEntangledState} and Eq. \eqref{EntangledState}. In both cases the operator performing the quantum reference frame transformation maps the Hilbert space of $B$ and $C$ ($\nu$) to the Hilbert space of $A$ and $B$, $S: \mathcal{H}^{(A)}_B \otimes \mathcal{H}^{(A)}_{C(\nu)} \rightarrow \mathcal{H}_A^{(C,\nu)} \otimes \mathcal{H}_B^{(C,\nu)}$, where $S$ is the translation operator $T$ in Eq. \eqref{EntangledState} and the boost operator $U$ in Eq. \eqref{NeutrinoEntangledState}. These operators can be more explicitly written as $T (-\hat{x}_C) = T_{AC}(-\hat{x}_C) \otimes T_B (-\hat{x}_C)$ and $U( \hat{\Lambda}_{\nu}) = U_{A\nu}(\hat{\Lambda}_{\nu}) \otimes U_B(\hat{\Lambda}_{\nu}) $. They take as input the state of $C (\nu)$ via the operators $\hat{x}_C$ and $\hat{\Lambda}_{\nu}$ and map the states of $B$ and $C (\nu)$ relative to $A$ to the states of $A$ and $B$ relative to $C (\nu)$ accordingly. The effect of the 'parity-swap operator' as defined in the refs. \cite{FQR2,FQR3,FQR4}, is encoded in $T_{AC}$ (respectively $U_{A\nu}$). As it was the case in Eq. \eqref{EntangledState}, the transformation to the neutrino frame brings about entanglement, which is easily seen from Eq. \eqref{NeutrinoEntangledState}.

\textbf{A gedankenexperiment} -- The previous considerations apply to all those particles which do not have a definite mass, and in particular to the neutral meson pairs $K^0 - \bar{K}^0, D^0 - \bar{D}^0, B^0 - \bar{B}^0$. The strangeness eigenstates are linear combinations of the mass eigenstates, that is, neglecting $CP$-violation $\ket{K^0 / \bar{K}^0} = \frac{1}{\sqrt{2}}\left( \ket{K_L} \pm \ket{K_S}\right)$. Only the latter carry a definite $4$-momentum, just like the neutrino mass states $\ket{p^{\mu}_j}_{m_j}$.
To show how the frame related entanglement may have phenomenological significance, consider the decay of a charged meson into three particles, such as \cite{PDG} $D^{+} \rightarrow \bar{K}^0 + e^+ + \nu_e $ or $B^+ \rightarrow \bar{D}^0 + l^+ + \nu_{l}$ for $l= e,\mu,\tau$. Consider first that the neutral meson has a definite $3$-momentum $\pmb{k}$ and that the state of the decay products, in the laboratory frame, is
\begin{eqnarray}
 \nonumber &&\ket{DP}^{LAB} = \\ \nonumber
 &&\frac{1}{\sqrt{2}}\left(\ket{(\pmb{k}, \omega_L)}_{m_L, K}^{LAB} - \ket{(\pmb{k}, \omega_S)}_{m_S, K}^{LAB} \right)\ket{p^{\mu}_{e^+}}_{e^+}^{LAB} \ket{p^{\mu}_{\nu_e}}_{\nu_e}^{LAB}  \ .
\end{eqnarray}
Here we are neglecting the oscillating nature of $\nu_e$ for simplicity. Consider now the same reaction, characterized by the same kinematical (Mandelstam) invariants $s,u,t$, but this time assume that the neutral meson momentum is very close to $0$ in the laboratory frame. Assume then that it is effectively possible to identify the rest frame of the neutral kaon with the laboratory frame. The state of the decay products in this frame is therefore obtained by a quantum reference frame transformation to the neutral meson rest frame, namely

\begin{eqnarray}\label{EntangledKaon}
 \nonumber \ket{DP}^{(K)}  &&=   \frac{1}{\sqrt{2}} \Big(\ket{\Lambda^{\mu}_{L,\nu}p^{\nu}_{e^+}}_{e^+}^{(K)} \ket{\Lambda^{\mu}_{L,\nu}p^{\nu}_{\nu_e}}_{\nu_e}^{(K)} \\ &&- \ket{\Lambda^{\mu}_{S,\nu}p^{\nu}_{e^+}}_{e^+}^{(K)} \ket{\Lambda^{\mu}_{S,\nu}p^{\nu}_{\nu_e}}_{\nu_e}^{(K)} \Big) \ ,
\end{eqnarray}

where we have omitted the (possibly dummy) state of the original laboratory $\ket{LAB}_{LAB}^{(K)}$. The Lorentz boosts $\Lambda_L, \Lambda_S$, assuming $\pmb{k}$ along the third axis, have the form of Eq. \eqref{ClassicalTransformation} as specified by the $4$-momenta of $K_L$ and $K_S$. We can see from Eq. \eqref{EntangledKaon} that the state of the leptons, previously a product state, becomes entangled in the new frame. This additional form of entanglement is exclusively related to the quantum reference frame transformation, and may or may not characterize the decay products of a reaction depending on the frame considered.

\textbf{Conclusions} -- We have discussed the role of quantum reference frames for mixed particles, showing  the necessity of quantum frame transformations to define the corresponding rest frame. We have pointed out how the frame-dependent entanglement may arise in physical processes involving mixed particles. In view of the fundamental role that the symmetry group of frame transformations plays in the classification of one particle states, and in the definition of elementary particles, the introduction of quantum frames may have a significant impact on quantum field theory. Our analysis reveals that a generalization in the sense of quantum frames is needed to properly describe mixed particles. These considerations may pave the way for a new quantum field theory in which the symmetry group itself has quantum properties.

\section*{Acknowledgements}
Partial financial support from MIUR and INFN is acknowledged.
A.C. also acknowledges the COST Action CA1511 Cosmology
and Astrophysics Network for Theoretical Advances and Training Actions (CANTATA).

\end{document}